\documentclass{article}
\usepackage{amscd,amssymb,bbb,amsmath}
\usepackage{algorithm}
\usepackage{algorithmic}
\usepackage{graphicx}
\usepackage{dsfont}

\newtheorem{theorem}{Theorem}[section]

\newtheorem{definition}{Definition}[section]

\usepackage{multirow}
\usepackage{dcolumn}
\usepackage[T1]{fontenc}
\usepackage[utf8]{inputenc}
\usepackage{tabularx,ragged2e,booktabs,caption}
\newcolumntype{C}[1]{>{\Centering}m{#1}}

\usepackage{tikz}
\usetikzlibrary{arrows.meta}
\usetikzlibrary{matrix, calc}

\usepackage{verbatim}
\usetikzlibrary{arrows,shapes}
\usetikzlibrary{hobby,backgrounds,calc,trees}

\title{Computational Analysis of the structural properties of Economic and Financial Networks}

\begin{document}

\pgfdeclarelayer{background}
\pgfsetlayers{background,main}

%% Adjacency matrix of graph
%% \  a  b  c  d  e  f  g
%% a  x  7     5
%% b  7  x  8  9  7
%% c     8  x     5
%% d  5  9     x 15  6
%% e     7  5 15  x  8  9
%% f           6  8  x 11
%% g              9  11 x

\tikzstyle{vertex}=[circle,fill=black!25,minimum size=20pt,inner sep=0pt]
\tikzstyle{selected vertex} = [vertex, fill=red!24]
\tikzstyle{edge} = [draw,thick,-]
\tikzstyle{weight} = [font=\small]
\tikzstyle{selected edge} = [draw,line width=5pt,-,red!50]
\tikzstyle{ignored edge} = [draw,line width=5pt,-,black!20]

\maketitle

\noindent
Frank Emmert-Streib$^{\mbox{a,b,*}}$, Aliyu Musa$^{\mbox{a,b,*}}$, Kestutis Baltakys$^{\mbox{c}}$,  Juho Kanniainen$^{\mbox{c}}$, Shailesh Tripathi$^{\mbox{a}}$, Olli Yli-Harja$^{\mbox{b,d}}$, Herbert Jodlbauer$^{\mbox{e}}$ and Matthias Dehmer$^{\mbox{e,f,g}}$

\vspace{0,2cm}
\noindent
$^{\mbox{a}}$
Predictive Medicine and Data Analytics Lab, Department of Signal Processing, Tampere University of Technology, Finland \\
*Corresponding author: v@bio-complexity.com \\

\noindent
$^{\mbox{b}}$
Institute of Biosciences and Medical Technology, Tampere, Finland\\

\noindent
$^{\mbox{c}}$
Industrial and Information Management, Tampere University of Technology, Tampere, Finland\\

\noindent
$^{\mbox{d}}$
Institute for Systems Biology, Seattle, USA\\

\noindent
$^{\mbox{e}}$
Institute for Intelligent Production, Faculty for Management, University of Applied Sciences Upper Austria, Steyr Campus, Austria\\

\noindent
$^{\mbox{f}}$
Department for Biomedical Computer Science and Mechatronics, UMIT - The Health and Lifesciences University, Hall in Tyrol, Austria\\

\noindent
$^{\mbox{g}}$
College of Computer and Control Engineering, Nankai University, Tianjin 300350, P. R. China\\

\begin{abstract}

In recent years, methods from network science are gaining rapidly interest in economics and finance. A reason for this is that in a globalized world the interconnectedness among economic and financial entities are crucial to understand and networks provide a natural framework for representing and studying such systems. In this paper, we are surveying the use of networks and network-based methods for studying economy related questions. We start with a brief overview of graph theory and basic definitions. Then we discuss descriptive network measures and network complexity measures for quantifying structural properties of economic networks. Finally, we discuss different network and tree structures as relevant for applications.

\end{abstract}

\section{Introduction}

In economics there is an increasing interest in recent years to investigate financial, economic, production and investment markets by means of networks. One reason for this interest is that a network, also called a graph, allows the convenient mathematical representation and analysis of a system with many interacting entities. This flexibility  is wide enough to accommodate all different types of economic networks existing, e.g., interbank networks, investment networks, director networks, ownership networks, financial networks, product networks and trade networks \cite{hochberg2007whom,arnold2006topology,boss_2004,degryse2004interbank,roukny_2014,vitali_2011,qui_2010,dhar2014prediction}. 

Despite the fact that the study of graphs and networks going back to Euler and Cayley \cite{euler_1736,cayley_1857} with a formalization of the theory, by {\it K\"{o}nig} in the 1930's \cite{Koenig_1936} and many interdisciplinary applications in mathematics  \cite{brandstaedt_1999,Diestel_2000,erdrenyi_1959,harary_1967}, computer science \cite{cormen_2001,even_1979}, physics \cite{harary_1967}, biology \cite{kauffman_1969,palsoon_2006,roberts_book,emmert_dehmer_microarray_wiley_book_2008} and sociology \cite{harary3,scott1,wasserman}, has a long lasting tradition the study of economic networks is lacking behind these other fields. One reason for this might be the difficulty in construction economic networks. For instance, while it is relatively easy to observe the acquaintanceships among a group of people or the molecular composition of chemicals leading to social and chemical networks the effect of one stock on another for constructing a financial network is considerably more difficult to infer requiring an appropriate statistical method and a dataset allowing to accomplish this. Fortunately, technological progress and the emergence of our digital society allow nowadays to tackle this problem.

Before we start with our review we want to mention a couple of possible applications. According to Hopp and Spearman \cite{hopp2011factory} a production system is a network of interacting parts where managing the interactions is as - or even more - important, than managing individual processes and entities. Graph theory is a powerful approach to model these interactions.

Historically, well-known methods are the program evaluation and review technique (PERT), developed by Kelly and Walker for the US Navy, and the critical path method (CPM) developed by Booz, Allen and Hamilton \cite{kelley1959critical}. Both methods help managers to schedule, monitor and control large as well as complex projects. In operations several complex projects may occur: optimal resource allocation for different flight rates of a space shuttle \cite{heileman1992simulation} or shutdown management and scheduling maintenance \cite{roberts1992shutdown}, to name only two. In \cite{hartmann2010survey,pinedo2012scheduling,kalinowski2014graph,morinaga2016facility} a good survey of the resource-constrained project scheduling problem is given.

Similar to project schedules in manufacturing, graphs can be used to describe complex precedence constraints and production schedules \cite{pinedo2012scheduling,kalinowski2014graph}. The broadly used Gantt charts in production planning have an equivalent graph representation. For flow shops with unlimited intermediate storage as well as limited intermediate storage a directed graph can be applied for the computation of different objectives like makespan, tardiness or number of tardy jobs. For job shop disjunctive graphs or bipartite graphs are suitable to model the minimization of the completion time. Assemble line balancing is a specific problem in production planning. Kilbridge and Wester developed a heuristic diagram of work elements based on precedence \cite{kilbridge1961heuristic}.

Network Location models \cite{amin2013multi} support the task to locate one or more new facilities in an existing network in order to minimize multi-objectives for instance some function of distance separating the new and existing facilities. Systematic Layout planning tries to find an optimal plant layout between technological limitations, organizational policies, safety considerations, space requirements, availability and product and process constraints by finding a maximally planar weighted graph \cite{morinaga2016facility}.

The purpose of this paper is to generate more awareness about the potential of a structural analysis of economic networks by reviewing approaches from graph theory and network science. We start in Section~(\ref{sec_graph_definitions}) by providing necessary preliminaries from graph theory. In Section (\ref{NA}), we review local and global descriptive network measures and in Section \ref{sec:network_complexity} we discuss measures that quantify the structural complexity of networks. Section \ref{NC} gives an overview of important network and tree classes as useful for the study of economic networks. This article finishes in Section (\ref{future}) and (\ref{con}) with a discussion of potential future directions and conclusions.

\section{Setting the framework from graph theory}\label{sec_graph_definitions}

Before we can begin surveying analysis methods of economic networks we need to provide the necessary preliminaries from graph theory. We start with basic definitions for undirected and directed graphs \cite{bang,harary}. 

\begin{definition}\label{simple_graph_undirected}
The pair $G=(V, E)$ where $V$ represents a finite set of vertices and $E$ the set of edges, $E \subseteq {V \choose 2}$ 
is called a finite undirected graph. 
\end{definition}

Throughout the paper, we set the cardinality of the vertex set $|V|:=N$. The cardinality of the edge set is denoted by $|E|$. In the following, we write $N(G)$ and $|E(G)|$ instead of $N$ and $|E|$ when it is necessary to emphasize that we refer to a specific graph $G$.

\begin{definition}\label{def_set}
$\mathcal{G}(N)$ denotes the set of undirected graphs having $N$ vertices.
\end{definition}

\begin{definition}\label{simple_graph_directed}
The pair $G=(V, E)$ where $V$ represents a finite set of vertices and $E$ the set of edges, $E \subseteq {V \times V}$, is called a finite directed graph.
\end{definition}

We emphasize that in this paper, we are only considering graphs with a finite vertex set. Hence, their edge set is also finite. For this reason, those graphs are called
finite graphs \cite{harary}. In contrast, infinite graphs possess a both infinite vertex set and an infinite edge set. They have been investigated, e.g., when studying growth models
for the world wide web, birth and death processes, random graph models, or to investigate mathematical symmetry by using Cayley graphs \cite{bollabas,chakrabarti2,harary,erdoes_1960}.

%
%\begin{definition}\label{teilgraph_def}
%Let $G=(V,E)$ be a finite graph.
%A graph $H:=(V',E')$ with $V' \subseteq V$ and $E' \subseteq E$ is called subgraph of
%$G$, $H \subseteq  G$. Moreover, $H$ is called an induced subgraph of $G$ if $E'$ contains all edges $e\in E$ that join vertices in $V'$.
%\end{definition}

We remark if $G=(V,E)$ is allowed to have loops (reflexive edges) and parallel edges, then $G$ is called a multigraph \cite{gross_2006,harary}.
In contrast, hypergraphs \cite{berge} are generalizations of the ordinary notation of a graph, we just introduced. Specifically, for an ordinary graph (see Def.~(\ref{simple_graph_undirected},~(\ref{simple_graph_directed})), an edge connects exactly two vertices whereas
a hyperedge can connect any number of vertices, see \cite{berge}.
Graphs which possess directed hyperedges are called directed hypergraphs and have been defined in \cite{gallo}.

A very important graph class are labeled graphs \cite{harary}. For instance, they have been used to model complex structures in various scientific disciplines
like biology \cite{felsenstein_2003,foulds_1992,semple_steel}, chemistry \cite{devillers_balaban_2000,Trinajstic_book}, sociology \cite{wasserman}, and mathematical psychology \cite{sommerfeld_1994,sommerfeld_habil}.

\begin{definition}\label{def_labelled_graphs}
%Let $G=(V,E) \in \mathcal{G}_{UC}$.
Let
\begin{equation}\label{def_vertex_alphabet}
A^G_V:= \{l^1_v,l^2_v,\ldots,l^{|A^G_V|}_v \},
\end{equation}
and
\begin{equation}\label{def_edge_alphabet}
A^G_E:= \{l^1_e,l^2_e,\ldots,l^{|A^G_E|}_e \},
\end{equation}
be unique (finite) vertex and edge alphabets, respectively.
$l_V: V \longrightarrow A^G_V$ and $l_E: E \longrightarrow A^G_E$ are
the corresponding edge and vertex labeling functions.
\begin{equation}
G:=(V,E, l_V, l_E),
\end{equation}
is called a finite, labeled graph.
\end{definition}

%The following definition describes if two graphs are topologically equivalent \cite{harary}.
%\begin{definition}
%Let $G=(V,E)$ and $H:=(V',E')$ be two finite graphs.
%We call $G$ and $H$ isomorphic ($G$ $\cong$ $H$) if there exists a
%bijective mapping $\phi: V \longrightarrow V'$ such that
%\begin{eqnarray}
%E \ni (v_i,v_j) :\Longleftrightarrow  (\phi(v_i),\phi(v_j)) \in E'.
%\end{eqnarray}
%The mapping $\phi$ is called isomorphism of $G$ on $H$.
%\end{definition}

For representing a graph or a network practically, the so-called adjacency matrix can be used \cite{harary}.
\begin{definition}
The adjacency matrix of a finite graph $G=(V,E)$ is defined by
\begin{equation}\label{def_adjazenz_matrix}
A_{ij}:=\left\{\begin{array}{ll}
                             1 & :  (v_i,v_j) \in E\\
                             0 & :  \mbox{otherwise}
                             \end{array} \right.
\end{equation}
\end{definition}

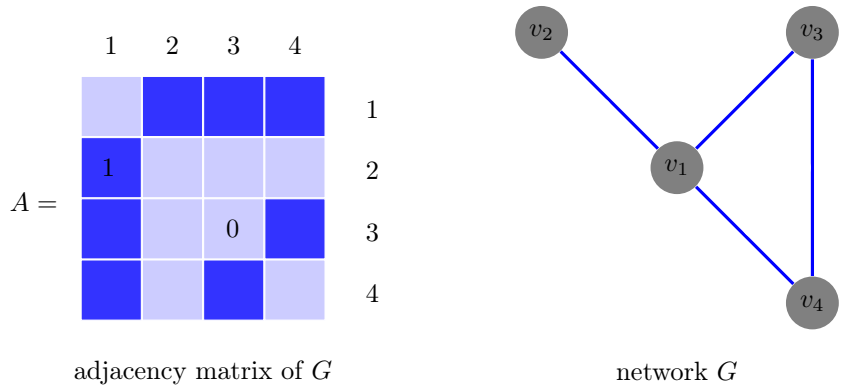
\begin{figure}
\centering
\begin{tikzpicture}[scale=1.8, auto,swap]

\matrix (input) [matrix of nodes,
                nodes={rectangle, draw=white, minimum size=.8cm}] at (-2.5,-1)
{
 1 & 2 & 3 & 4 \\
 |[fill=blue!20]| & |[fill=blue!80]|  & |[fill=blue!80]|   & |[fill=blue!80]|      \\
 |[fill=blue!80]| & |[fill=blue!20]|  & |[fill=blue!20]|  & |[fill=blue!20]|   \\
 |[fill=blue!80]| & |[fill=blue!20]|  & |[fill=blue!20]|   & |[fill=blue!80]|      \\
 |[fill=blue!80]| & |[fill=blue!20]|  & |[fill=blue!80]|   & |[fill=blue!20]|      \\
};
\node [below=8pt] at (input.south) {adjacency matrix of $G$};

\matrix (input) [matrix of nodes,
                nodes={rectangle, draw=white, minimum size=.8cm}] at (-1.25,-1.25)
{
 1 \\
 2 \\
 3\\
 4\\
};

    % Draw a 7,11 network
    % First we draw the vertices
            \node[vertex, fill=gray] (x1) at (0,0) {$v_2$};
            \node[vertex, fill=gray] (x2) at (2,0) {$v_3$};           
            \node[vertex, fill=gray] (x3) at (2,-2) {$v_4$};
            \node[vertex, fill=gray] (x4) at (1,-1) {$v_1$};
                    
    % Connect vertices with edges and draw weights

    \begin{scope}[>={Stealth[black]},
              every node/.style={fill=white,circle},
              every edge/.style={draw=blue,very thick}]
    \path [-] (x2) edge[draw=blue] node {} (x3);
    \path [-] (x1) edge node {} (x4);
    \path [-] (x2) edge node {} (x4);
    \path [-] (x3) edge node {} (x4);

\end{scope}

\node[] (x9) at (-3.75,-1.25) {$A=$};
\node[] (x10) at (-3.2,-1.0) {$1$};
\node[] (x11) at (-2.28,-1.45) {$0$};
\node[] (x12) at (1,-2.5) {network $G$};
    
\end{tikzpicture}
\caption{Representation of a network $G$ by an adjacency matrix $A$. Due to the undirectedness of $G$ the matrix $A$ is symmetrical. }\label{fig.adj}
\end{figure}
In Fig. \ref{fig.adj} we show an example for an adjacency matrix $A$ of an undirected network $G$ shown on the right hand side. This this case the matrix $A$ is symmetric. 

Based on the adjacency matrix of a network, eigenvalues \cite{cvetkovic1,harary} can be defined. 
\begin{definition}
The spectrum of $G$ consists of the sets
$M_{\lambda}=\{\lambda_1, \lambda_2, \ldots, \lambda_k \}$ and $M=\{n_1, n_2, \ldots, n_k \}$
where $n_i$ denotes the multiplicity of the zero $\lambda_i$ of the equation $\det(A-\lambda U) =0$, the characteristic polynomial of $G$ where $A$ is its adjacency matrix and $U$ is the unit matrix. 
\end{definition}
Examples for eigenvalue-based models that have been applied in the context of economic networks can be found in \cite{koenig_2009}.

\section{Descriptive Network Measures}\label{NA}

In the following sections we present quantitative network measures allowing to perform a descriptive network analysis. Many of the measures have their origin in the social sciences, chemistry or information sciences \cite{allen_2002,bonchev_2,bonchev_book_2005,wasserman}. If not stated otherwise, we are assuming that the networks have undirected edges.

\subsection{Node Degree and Degree distribution}

The degree $k_i$ of a vertex $i$ is the number of edges which are incident with vertex $i$ given by
\begin{eqnarray}
k_i= \sum_j A_{v_i,v_j}
\end{eqnarray}
From this the degree distribution \cite{bornholdt,mason_2007} is obtained by
\begin{equation}\label{degree_distribution_eq}
P(k) := \frac{\delta_k}{N},
\end{equation}
where $\delta_k$ denotes the number of vertices in the network $G$ having a degree of $k$ and $N$ is the total number of nodes. Eqn.~(\ref{degree_distribution_eq}) corresponds to the proportion of vertices in $G$ having a degree of $k$. Formally, $\delta_k$ can be written as,
\begin{eqnarray}
\delta_k = \sum_{i=1}^N I(k_i = k)
\end{eqnarray}
where $I()$ is the indicator function giving $1$ for a true argument and $0$ otherwise. Another meaning of Eqn.~(\ref{degree_distribution_eq}) is that a randomly chosen vertex in the network has a probability of $P(k)$ to be linked with $k$ other vertices.

It was an interesting and important finding that many real world networks like the World Wide Web (WWW), the Internet, social networks, citation networks or food webs \cite{adamic1,bornholdt,brandes_2005} are not Poison distributed like random networks (see Sec. \ref{design_and_random_graphs_sec} for a detailed discussion of random networks) but follow a power law distribution, i.e.,
\begin{equation}\label{powerlaw_eq}
P(k)  \sim k^{-\gamma}, \quad \gamma >1.
\end{equation}

In contrast to the above measures characterizing properties of individual nodes, there are also measures for characterizing the whole network. For instance, the average degree for the entire network is:
\begin{equation}
k = k(G):=\sum_{v\in V} \frac{k_v}{N}, \label{mdeg}
\end{equation}
Finally, the edge density of $G$ is defined by
\begin{equation}\label{edge_density_eq}
\beta(G):= \frac{|E|}{ {N \choose 2} }.
\end{equation}
Here the denominator gives the total number of possible edges for a network with $N$ nodes, which corresponds to a fully connected network. Further network statistics and advanced aspects can be found in, e.g.,
\cite{brinkmeier_2005,skorobogatov}.

\subsection{Clustering coefficient}

\begin{figure}
\centering
\begin{tikzpicture}[scale=1.8, auto,swap]
    % Draw a 7,11 network
    % First we draw the vertices
            \node[vertex, fill=gray] (x1) at (0,0) {$$};
            \node[vertex, fill=gray] (x2) at (2,0) {$$};
                        
            \node[vertex, fill=gray] (x3) at (-0.75,-1) {$$};
            \node[vertex, fill=orange] (x4) at (1,-1) {$v_i$};
            \node[vertex, fill=gray] (x5) at (0,-2) {$$};            
           
            \node[vertex, fill=gray] (x6) at (2,-2) {$$};
          
    % Connect vertices with edges and draw weights

    \begin{scope}[>={Stealth[black]},
              every node/.style={fill=white,circle},
              every edge/.style={draw=red,very thick}]
    \path [-] (x1) edge[draw=blue] node {} (x3);
    \path [-] (x1) edge node {} (x4);
    \path [-] (x2) edge node {} (x4);
    \path [-] (x5) edge node {} (x4);
     \path [-] (x6) edge node {} (x4);
     \path [-] (x3) edge[draw=blue] node {} (x5);
     
       \path [-] (x1) edge[bend right=20, draw=blue] node {} (x5); 
       \path [-] (x3) edge node {} (x4);
   
\end{scope}

\node[] (x9) at (2.5,-1) {$e_i=3, n_i=5$};
    
\end{tikzpicture}
\caption{An example for a clustering coefficient. The node $v_i$ has $n_i  =5$ connections and $e_i  =3$. This results in $C_i = 3/10$. }\label{fig.cluster}
\end{figure}
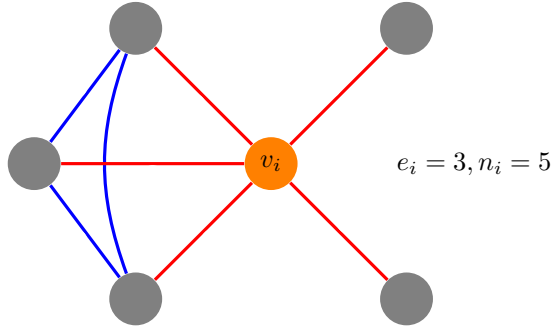

The clustering coefficient $C_i$ is a local measure \cite{watts2} defined for every vertex $i$. For an undirected network it is defined by
\begin{eqnarray}\label{eq_clustering_coefficient}
C_i = \frac{2e_i}{n_i(n_i-1)},
\end{eqnarray}
where $n_i$ is the number of neighbors of vertex $i$ and $e_i$ is the number of adjacent pairs between all neighbors of $i$ \cite{watts2}. In Fig. \ref{fig.cluster} we show an example.

\subsection{Path-based measures}\label{sec:pbm}

The next type of measures involve more than one node for their calculation, instead, they are based on the path between nodes \cite{brinkmeier_2005,buckley_harary,halin,skorobogatov}.

Let $G=(V,E)$ be a connected graph and
\begin{equation}
(d(v_i,v_j))_{v_i,v_j \in V},
\end{equation}
be the distance matrix, where $d(v_i,v_j)$ denotes the distance (length of a shortest path) between the nodes $v_i$ and $v_j$. From this the average distance of a network follows by
\begin{equation}\label{averagedistance_eq}
\bar{d}(G):= \frac{1}{ {N \choose 2}  }\sum_{1\le i<j\le N} d(v_i,v_j).
\end{equation}

Additional graph metrics \cite{skorobogatov} based on the distance matrix are
\begin{equation}\label{eccen_eq}
\sigma(v)= \max_{u\in V}d(u,v),
\end{equation}
\begin{equation}\label{diameter_eq}
\rho(G)= \max_{v\in V}\sigma(v),
\end{equation}
and
\begin{equation}\label{radius_eq}
r(G)= \min_{v\in V}\sigma(v).
\end{equation}
The above entity $\sigma(v)$ is called the eccentricity of $v \in V$, $\rho(G)$ is the diameter of $G$, and $r(G)$ is the radius of the graph.

\subsection{Network Centrality Measures: Identifying important nodes}

There is a large family of measures called centrality measures \cite{freeman1977} that have their origin in the social sciences \cite{hage_harary_1995,wasserman}. The goal of these measures has been to identify nodes in a networks that are important in terms of communication. 

  Conceptually, one distinguishes between two fundamentally different types of centrality measures \cite{freeman1977,freeman1979}. The first type is called {\it point centrality} measures and the second {\it graph centrality} measures. The difference is that the former characterize local properties of a graph whereas the latter ones characterize global properties.

For an undirected graph $G=(V,E)$, the degree centrality of a vertex $v\in V$ is
defined as its degree, i.e.,
\begin{equation}\label{degree_centrality_eq}
C_D(v)= k_v.
\end{equation}

\begin{figure}
\centering
\begin{tikzpicture}[scale=1.8, auto,swap]
    % Draw a 7,11 network
    % First we draw the vertices
            \node[vertex, fill=gray] (x1) at (-2,0) {$v_i$};
            \node[vertex, fill=orange] (x2) at (0,0) {$v_k$};        
            \node[vertex, fill=gray] (x3) at (2,0) {$v_j$};

    % Connect vertices with edges and draw weights

    \begin{scope}[>={Stealth[black]},
              every node/.style={fill=white,circle},
              every edge/.style={draw=red, very thin}]
    %\path [-] (x1) edge[draw=blue] node {} (x3); 
     
      % \path [-] (x1) edge[bend right=20, draw=blue] node {} (x5); 
       
        \draw [blue] (x1) % Draws a line
      to [out=270,in=180] (0,-2) 
      to [out=0,in=270] (x3);   
       
       \draw [blue] (x1) % Draws a line
      to [out=90,in=180] (0,2) 
      to [out=0,in=90] (x3);   
      
       \draw [blue] (x1) % Draws a line
      to [out=90,in=180] (0,1.65) 
      to [out=0,in=90] (x3);   
      
       \draw [green] (x1) % Draws a line
      to [out=10,in=180] (0,0) 
      to [out=0,in=190] (x3);   
       \draw [green] (x1) % Draws a line
      to [out=10,in=135] (0,0) 
      to [out=-35,in=190] (x3);   
       \draw [green] (x1) % Draws a line
      to [out=-10,in=180] (0,0) 
      to [out=0,in=170] (x3);   
       \draw [green] (x1) % Draws a line
      to [out=-10,in=225] (0,0) 
      to [out=45,in=170] (x3);   
      
  \draw[color=blue] (x1) to [bend left=30] (x3);
  \draw[color=blue] (x1) to [bend right=30] (x3);
  \draw[color=blue] (x1) to [bend left=50] (x3);
  \draw[color=blue] (x1) to [bend right=50] (x3);
  \draw[color=blue] (x1) to [bend left=70] (x3);
  \draw[color=blue] (x1) to [bend right=70] (x3);
\end{scope}

\node[] (x9) at (3.0,-1.0) {
\begin{tabular}{l}
$\sigma_{v_i, v_j}=13$  \\
$\sigma_{v_i, v_j}(v_k)=4$ \\
\end{tabular}
};
\node[vertex, fill=orange] (x2) at (0,0) {$v_k$}; 
    
\node[vertex, fill=gray, scale=0.5] (x11) at (-1.5,1.4) {$$};%ggg
\node[vertex, fill=gray, scale=0.5] (x11) at (1.5,-1.4) {$$};
\node[vertex, fill=gray, scale=0.5] (x11) at (1.5,1.4) {$$};
\node[vertex, fill=gray, scale=0.5] (x11) at (-1.0,-0.8) {$$};
\node[vertex, fill=gray, scale=0.5] (x11) at (-1.5,-0.8) {$$};
\node[vertex, fill=gray, scale=0.5] (x11) at (1.5,-0.8) {$$};
\node[vertex, fill=gray, scale=0.5] (x11) at (1.5,0.8) {$$};
\node[vertex, fill=gray, scale=0.5] (x11) at (0.4,-0.6) {$$};
\node[vertex, fill=gray, scale=0.5] (x11) at (-0.4,-0.6) {$$};
\node[vertex, fill=gray, scale=0.5] (x11) at (0.4,0.6) {$$};
\node[vertex, fill=gray, scale=0.5] (x11) at (-0.3,1.63) {$$};
    
\end{tikzpicture}
\caption{Visualization of the betweenness centrality measure. The gray nodes are further ndoes in the network.}\label{fig.between}
\end{figure}
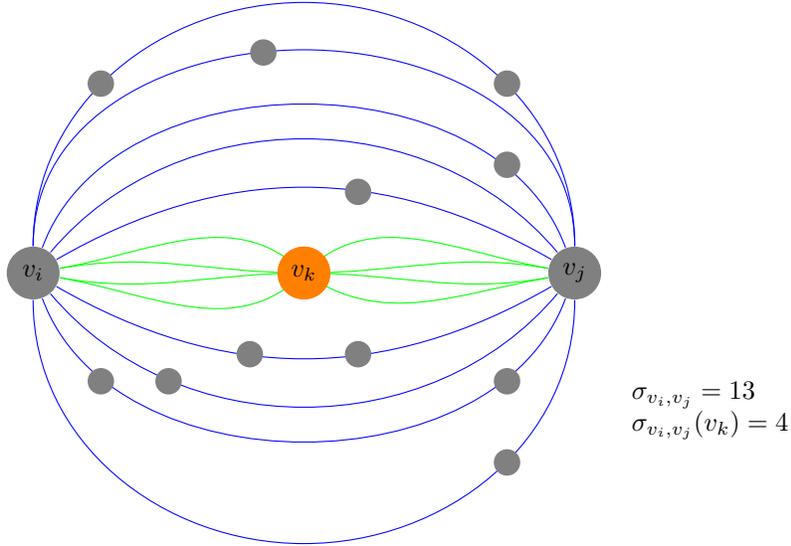

The next measure $C_B(v_k)$ is called {\it betweenness} centrality,
\begin{eqnarray}
C_B(v_k) &=& \sum_{v_i, v_j \in V, v_i \ne v_j} \frac{\sigma_{v_iv_j}(v_k)}{\sigma_{v_iv_j}},  \label{eq_vertex_centrality_4}
\end{eqnarray}
$C_B(v_k)$ is based on distances, see, e.g., \cite{freeman1977,sabidussi_1966,scott1}. Here $\sigma_{v_iv_j}$ stands for the number of shortest paths from $v_i$ to $v_j$ and $\sigma_{v_iv_j}(v_k)$ for the number of shortest paths from $v_i$ to $v_j$ that include $v_k$. Thus, the quantity
\begin{eqnarray}
\frac{\sigma_{v_iv_j}(v_k)}{\sigma_{v_iv_j}},
\end{eqnarray}
can be interpreted as the probability that $v_k$ lies on a shortest path connecting $v_i$ with $v_j$.
Consequently, $C_B(v_k)$ determines the appearance of $v_k$ on all shortest paths in the corresponding network. In Fig. \ref{fig.between} we show a visualization of $C_B(v_k)$.

Another well-known measure is the centrality index called {\it closeness},
\begin{eqnarray}
C_C(v_k) &=& \frac{1}{\sum_{i=1}^{N}d(v_k,v_i)}.  \label{eq_vertex_centrality_1}
\end{eqnarray}
$d(v_k,v_i)$ denotes the number of edges on a shortest path between $v_k$ and $v_i$. In case there are multiple shortest paths connecting $v_k$ with $v_i$, $d(v_k,v_i)$ remains unchanged.
Note that $C_C(v_k)$ can be used to evaluate
how close is a vertex to other vertices in a given network.

%This is done in terms of communication distance as measured by the number of edges between two vertices if connected via the shortest path.

The previously mentioned measures are local centrality measures because they determine the centrality of a single vertex within a network.
In contrast, we now present the definition of a global measure called {\it graph centrality}.
%
%measures have been suggested which form extensions to the three measures discussed above \cite{freeman1979}.
%
Here, the crucial idea is to use these individual measures to obtain an average characteristic for the whole network:
\begin{eqnarray}
C_x = \frac{ \sum_{i=1}^N \big(C_x(v^m) - C_x(v_i) \big)}{ C_x^{max}}.
\end{eqnarray}
$x$ denotes any of the three point (local) centrality measures:
\begin{eqnarray}
C_x(v^m) = \max_i \{ C_x(v_i) \}.
\end{eqnarray}
$C_x(v^m)$ is the maximum of $C_x(v_i)$ determined for the given network and $C_x^{max}$ denotes the maximal value possible for $G \in \mathcal{G}(N)$ (see Definition~(\ref{def_set})),
\begin{eqnarray}
C_x^{max} = \max_{G \in \mathcal{G}(N)} \sum_{i=1}^N \big( C_x(v^m) - C_x(v_i) \big).
\end{eqnarray}
As special graph centrality measures, we obtain exemplarily \cite{brandes_2005,freeman1977,wasserman},
\begin{eqnarray}
C_d = \frac{ \sum_{i=1}^N C_d(v^m) - C_d(v_i)}{ N^2 - 3N + 2}, \label{gdc}
\end{eqnarray}
\begin{eqnarray}
C_b = \frac{ \sum_{i=1}^N C_d(v^m) - C_d(v_i)}{ N^3 - 4N^2 + 5N - 2},
\end{eqnarray}
%For the betweennes centrality of a network \cite{freeman1977} one obtains
%
and
\begin{eqnarray}
C_c = \frac{2N-3 }{ N^3 - 4N^2 + 5N - 2} \sum_{i=1}^N \big( C_d(v^m) - C_d(v_i) \big).
\end{eqnarray}
%For the closeness centrality of a network one obtains
%
%Intuitively, the denominator is obtained by remembering that the maximal degree of a vertex is $N-1$, hence, $C_x(v^m) - C_x(v_i) = N-2$ because the minimal degree is one.
%This number multiplied by $N-1$ (one vertex needs to have degree one) gives the denominator in (\ref{gdc}).
Further details and applications of these measures can be found in \cite{brandes_2005,freeman1977,wasserman}.

Aside from the measures presented so far, which are classic centrality measures there are several extended measures. For instance,  Bonacich \cite{bonacich1972} introduced the eigenvector centrality,
\begin{equation}
C_e = x^{max} = \frac{1}{\lambda^{max}} Ax^{max} .
\end{equation}
The idea of this measure is
to express that an important vertex is connected to important neighbors. For calculating $C_e$, one needs to determine the eigenvector of
the underlying adjacency matrix $A$ of a graph $G$ corresponding to the largest eigenvalue.
Let's assume $\lambda^{max}$ denotes this largest eigenvalue and $x^{max}$ the corresponding eigenvector.
It is important to note that $C_e$ is a point centrality measure because each vertex in the network obtains a value corresponding
to the component of $C_e$.  Further eigenvector centrality measures have been investigated in \cite{koschuetzki_2005}.

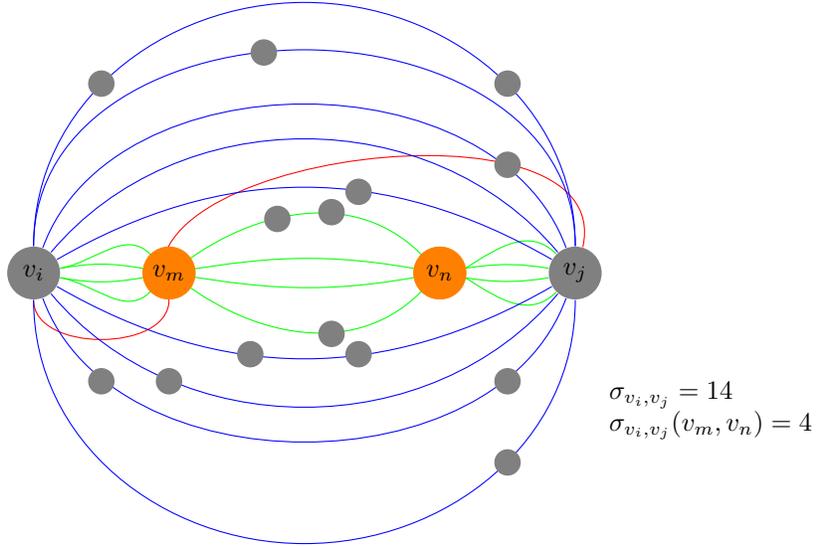
\begin{figure}
\centering
\begin{tikzpicture}[scale=1.8, auto,swap]
    % Draw a 7,11 network
    % First we draw the vertices
            \node[vertex, fill=gray] (x1) at (-2,0) {$v_i$};
            \node[vertex, fill=orange] (x2) at (-1,0) {$v_m$};        
            \node[vertex, fill=gray] (x3) at (2,0) {$v_j$};
            \node[vertex, fill=orange] (x4) at (1,0) {$v_n$};

    % Connect vertices with edges and draw weights

    \begin{scope}[>={Stealth[black]},
              every node/.style={fill=white,circle},
              every edge/.style={draw=red, very thin}]
    %\path [-] (x1) edge[draw=blue] node {} (x3); 
     
      % \path [-] (x1) edge[bend right=20, draw=blue] node {} (x5); 
        \draw [red] (x1) % Draws a line
      to [out=180,in=90,bend right=90] (x2) 
      to [out=270,in=0,bend left=110] (x3);

        \draw [blue] (x1) % Draws a line
      to [out=270,in=180] (0,-2) 
      to [out=0,in=270] (x3);   
       
       \draw [blue] (x1) % Draws a line
      to [out=90,in=180] (0,2) 
      to [out=0,in=90] (x3);   
      
       \draw [blue] (x1) % Draws a line
      to [out=90,in=180] (0,1.65) 
      to [out=0,in=90] (x3);   
      
       \draw [green] (x1) % Draws a line
      to [out=10,in=170] (x2) 
      to [out=-10,in=190] (x4) 
      to [out=10,in=170] (x3);   
      
       \draw [green] (x1) % Draws a line
      to [out=-10,in=225] (x2) 
to [out=45,in=135] (x4)
      to [out=-20,in=225] (x3);   
      
       \draw [green] (x1) % Draws a line
      to [out=10,in=135] (x2) 
to [out=-45,in=225] (x4)
      to [out=20,in=135] (x3);   
      
       \draw [green] (x1) % Draws a line
      to [out=-10,in=190] (x2) 
      to [out=10,in=170] (x4)
      to [out=-10,in=190] (x3);   
      
%       \draw [green] (x1) % Draws a line
%      to [out=-10,in=225] (x2) 
%      to [out=45,in=170] (x3);   
      
  \draw[color=blue] (x1) to [bend left=30] (x3);
  \draw[color=blue] (x1) to [bend right=30] (x3);
  \draw[color=blue] (x1) to [bend left=50] (x3);
  \draw[color=blue] (x1) to [bend right=50] (x3);
  \draw[color=blue] (x1) to [bend left=70] (x3);
  \draw[color=blue] (x1) to [bend right=70] (x3);
\end{scope}

\node[] (x9) at (3.0,-1.0) {
\begin{tabular}{l}
$\sigma_{v_i, v_j}=14$  \\
$\sigma_{v_i, v_j}(v_m, v_n)=4$ \\
\end{tabular}
};
\node[vertex, fill=orange] (x2) at (-1,0) {$v_m$}; 
\node[vertex, fill=orange] (x4) at (1,0) {$v_n$}; 
    
\node[vertex, fill=gray, scale=0.5] (x11) at (-1.5,1.4) {$$};%ggg
\node[vertex, fill=gray, scale=0.5] (x11) at (1.5,-1.4) {$$};
\node[vertex, fill=gray, scale=0.5] (x11) at (1.5,1.4) {$$};
\node[vertex, fill=gray, scale=0.5] (x11) at (-1.0,-0.8) {$$};
\node[vertex, fill=gray, scale=0.5] (x11) at (-1.5,-0.8) {$$};
\node[vertex, fill=gray, scale=0.5] (x11) at (1.5,-0.8) {$$};
\node[vertex, fill=gray, scale=0.5] (x11) at (1.5,0.8) {$$};
\node[vertex, fill=gray, scale=0.5] (x11) at (0.4,-0.6) {$$};
\node[vertex, fill=gray, scale=0.5] (x11) at (-0.4,-0.6) {$$};
\node[vertex, fill=gray, scale=0.5] (x11) at (0.4,0.6) {$$};
\node[vertex, fill=gray, scale=0.5] (x11) at (-0.3,1.63) {$$};

\node[vertex, fill=gray, scale=0.5] (x11) at (-0.2,0.4) {$$};
\node[vertex, fill=gray, scale=0.5] (x11) at (0.2,0.45) {$$};
\node[vertex, fill=gray, scale=0.5] (x11) at (0.2,-0.45) {$$};
    
\end{tikzpicture}
\caption{Visualization of the joint betweenness centrality measure. The gray nodes are further ndoes in the network that are different to $v_m$ and $v_n$.}\label{fig.jbetween}
\end{figure}

A conceptual extension of betweenness centrality \cite{shuja} has been provided by joint betweenness centrality (JBC) \cite{emmert2007functional}. JBC is a non-local measure because it quantifies the number of paths that flow through pairs of nodes in a network. This
centrality measure is defined by
\begin{eqnarray}
C_{jb}(v_m,v_n) = \sum_{v_i,v_j \in V, v_i \ne v_j}  \frac{ \sigma_{v_i,v_j}(v_m,v_n)}{\sigma_{v_i,v_j}}, \label{def:JBC}
\end{eqnarray}
JBC is evaluating the joint occurrence of two vertices on shortest communication paths in the network. Here $\sigma_{v_i,v_j}$ gives the number of shortest paths connecting $v_i$ with $v_j$ and $\sigma_{v_i,v_j}(v_m,v_n)$ gives the number of shortest paths connecting $v_i$ with $v_j$ that contain the vertices $v_m$ and $v_n$. In Fig. \ref{fig.jbetween} we show a visualization of the joint betweenness centrality measure. Further application examples of betweenness centrality and other variants in the context of economy can be found in Shuja \cite{shuja}. 

For the general application of centrality measures normalizations have been found to be useful. For instance, the analysis in \cite{emmert2007functional} used the following normalization,
\begin{eqnarray}
C_{jb}(v_m,v_n) =  \sum_{v_i,v_j \in V, v_i \ne v_j}  \frac{ \sigma_{v_i,v_j}(v_m,v_n) }{\sigma_{max}}, \label{def:JBCs}
\end{eqnarray}
where
\begin{eqnarray}
\sigma_{max} = \max_{v_i,v_j} \{ \sigma_{v_i,v_j} \}.
\end{eqnarray}

Examples of economic networks that have been analyzed by applying network centrality measures are prevalent. For instance, network centrality measures have been studied to identify systemically important financial institutions of the Turkish interbank market \cite{KUZUBAS2014203}. Specifically, the authors investigated the main borrower role of Demirbank in the crash of the banking system in 2000. A similar study of interbank networks can be found in \cite{temizsoy2016network} where data from the e-MID market in the Euro Area and US have been analyzed.  Another study using centrality measures can be found in \cite{sharma2017financial}. Their study aimed to identify core economic sectors of $20$ countries worldwide providing a linkage between financial networks and the underlying economic fundamentals. For their analysis they utilized eigenvector centrality. For further application examples to economic networks please see \cite{giudici2016graphical,AAAFR7,vitali_2014}.

\section{Network Complexity}\label{sec:network_complexity}

In contrast to the quantitative measures discussed so far, network complexity measures, which will be discussed in this section, evaluate the network as a whole. Here the term {\it complexity} is in general broadly defined but refers often to the well-known Kolmogorov complexity \cite{kolmogorov_1965,li_1997}. The underlying idea of network complexity measures is to assess the structural complexity as expressed by the intricate linking or branching structure of a graph.

In general, economic networks can be represented by undirected or directed networks. However, both types are topological networks which are amenable to a structural analysis in form of network complexity. The network complexity measures we are going to discuss in this section are quantitative measures. That means, an economic network will be mapped to a real number for determining the complexity thereof. This value can be seen as an {\it index} characterizing the network.

\subsection{Network Complexity based on Information Theory}\label{sec:network_complexity_info}

An important class of network complexity measures that is of relevance for practical applications is based on information theory. Information-theoretic complexity measures have been applied to many scientific areas such as biology, computer science or chemistry  \cite{bonchev_2,emmert_streib_dehmer_entropy,dehmer_CBAC_2007,mehler_2009}. In the following we review some of the most important measures from this area. 

We start with a network  $G$, where $X$  is  a graph invariant, and $\tau$ is an equivalence criterion. This leads to distributions like \cite{bonchev_2}:
\begin{equation}\label{eq_scheme_prob}
\begin{pmatrix} 1 & 2 & \cdots & k \\ |X_1| & |X_2|& \cdots & |X_k| \\ p_1 & p_2 & \cdots & p_k  \end{pmatrix} .
\end{equation}
The first row stands for the  equivalency classes and the second row are the cardinalities of the obtained partitions. 
From this we calculate probabilities by  $p_i=\frac{|X_i|}{|X|}$, for each partition corresponding to the third row of the matrix. That means
$\mathcal{P}_G=(p_1,\ldots,p_k)$ represents a probability distribution of $G$. 
By using the well-known Shannon-entropy \cite{shannon_weaver}, one obtains 
\begin{align}
I(G,\tau) & =  |X| \log(|X|) - \sum_{i=1}^{k} |X_i| \log(|X_i|), \label{eq1} \\
\bar{I}(G,\tau) & =  - \sum_{i=1}^{k} \frac{|X_i|}{|X|} \log\left( \frac{|X_i|}{|X|} \right). \label{eq2}
\end{align}
Eqn.~(\ref{eq1}) stands for the so-called  total information content of $G$  whereas Eqn.~(\ref{eq2}) is the  mean information content.
That means, once we have a given economic network, these two measures can be computed straight forward. 

Another method to determine the entropy of econmic networks is due to Dehmer  \cite{dehmer_cby_2007}. 
Instead of determining partitions by using a graph invariant $X$, we assign
a probability value to each vertex of a network. We do this by using an information functional $f$ 
that captures structural information of $G$. So, if we apply Shannons-entropy again, we obtain
\begin{equation}\label{eq_structural_information_content_general}
I_{f}(G):=-\sum_{i=1}^{N}{\frac{f(v_i)}{\sum_{j=1}^{N}f(v_j)} \log\left(\frac{f(v_i)}{\sum_{j=1}^{N}f(v_j)}\right)}.
\end{equation}
Once we choose concrete information functionals, we obtain concrete graph entropy measures. Examples are information functions based on  metrical properties of graphs namely \cite{dehmer_cby_2007,dehmer_varmuza_2009}:
\begin{multline}
f^{V_1}(v_i):= {\alpha}^{c_1|S_1(v_i,G)| + c_2|S_2(v_i,G)| + \cdots + c_{\rho(G)}|S_{\rho(G)}(v_i,G)| }, \\
c_k >0, 1 \leq k \leq \rho(G), \alpha >0.
\end{multline}

Note that the  parameters $c_k>0$ can be used to weight structural characteristics or differences of $G$ in each sphere. They need to  be chosen
such that they are all different, e.g., $c_1 > c_2 > \cdots > c_{\rho(G)}$, see \cite{dehmer_CBAC_2007}. Its evident that the choice of the  $c_k>0$ has an impact on the resulting measured value. For special economic networks posssing special topological properties like pathness or a large number of cycles, these parameters could be learned  systematically. 
When applying graph entropy measures to hierarchical economic networks representing hierarchical business group graphs, 
Altomonte and Rungi \cite{altomonte_2013} generalized the work of \cite{em_j15}.  They defined a new measure called 'Group
Index of Complexity' (GIC) which is given by
\begin{eqnarray}
GIC(G) = \sum_l^L l \frac{n_l}{N} \log \Big( \frac{N}{n_l} \Big)  
\end{eqnarray}
Here $L$ is the number of hierarchy levels, $n_l$ is the number of affiliates on hierarchical level $l$ and $N$ is the total number of affiliates in that network.
Altomonte and Rungi \cite{altomonte_2013} found a negative correlation between a vertical integration and the hierarchical
complexity of business groups. Further, they also determined a 
positive correlation between the hierarchical complexity of a business group and their productivity.

The last contribution we mention in this section is due to Bekiros et al.  \cite{Bekiros_2017}. They used information-theoretic quantities to measure the centrality
of economic networks. An important result of  \cite{Bekiros_2017} is that the authors proved evidence of the disparity of correlation and entropy-based centrality measurements for all markets between pre- and post-crisis periods.

\section{Different types of Networks and Trees}\label{NC}

Networks models in general have been useful for many reasons. In fact, networks enable an immediate visualization of complex interrelations among important 
players of a system under consideration. Also, networks constitute a mathematical representation which can be analyzed rigorously. 
We start this section by discussing important network classes.

\subsection{Random Networks}\label{design_and_random_graphs_sec}

Random networks are the first type of networks we have been studied extensively. For instance, 
the seminal work of Erd\"os and R\'{e}nyi \cite{erdrenyi_1959,erdrenyi_1960} started this development. 
Putting it simply, a random graph with $N$ vertices can be obtained by connecting every pair of vertices with a probability of $p$. 
The expected number of edges for a (undirected) network constructed this way is
\begin{eqnarray}
E(n) = p\frac{N(N-1)}{2}.
\end{eqnarray}

%\subsubsection{Degree distribution}
The degree distribution of a vertex $i$ in a random network follows a binomial
\begin{eqnarray}
P(k_i = k) = {N-1 \choose k} p^k (1-p)^{N-1-k}, \label{degbin}
\end{eqnarray}
because the maximal degree of vertex $i$ is at most $N-1$, the probability that the vertex has $k$ edges is $p^k (1-p)^{N-1-k}$ and there are ${N-1 \choose k}$ possibilities to choose $k$ edges from $N-1$ vertices.
By going to the limit $N \rightarrow \infty$, Eqn.~(\ref{degbin}) gives
\begin{eqnarray}
P(k_i = k) \sim \frac{z^k \exp(-z)}{k!}.
\end{eqnarray}
Here $z = p(N-1)$ is the expected number of edges for a vertex. That means for large $N$, the degree distribution of a vertex in a random network can be approximated by the Poison distribution. For this reason random networks are also called Poison random networks \cite{newman_2003}.

Further one can show that the degree distribution of a random network (instead of just a vertex) follows also approximately a Poison distribution
\begin{eqnarray}
P(X_k = r) \sim \frac{z^r \exp(-z)}{r!}.
\end{eqnarray}
It can be interpreted as there are $X_k=r$ vertices in the network that possess degree $k$ \cite{AlbertBara_2002rmp}.

For random networks the clustering coefficient $C_i$ of a vertex $i$, see Eqn.~(\ref{eq_clustering_coefficient}), assumes a very simple value. Specifically, because the  average degree of a vertex can be approximated by $z=p(N-1)\sim pN$, it follows 
\begin{eqnarray}
C_i \sim \frac{z}{N} =p ,
\end{eqnarray}

\subsection{Trees}\label{sec:tree}

A simple but non-trivial network is a tree. \cite{harary}. In general, a {\it tree} is connected and ayclic \cite{harary}. 
We state a theorem showing that a tree can be characterized by various properties, see \cite{ihringer}.

\begin{theorem}
Let $G$ be a graph having $N$ vertices. Then, the following assertions are equivalent:
\begin{enumerate}
\item $G=(V,E)$ is a tree.
\item Every two vertices of $G$ are connected by a unique path.
\item $G$ is connected, but for every edge $e \in E$ is $G\backslash \{e\}$ disconnected.
\item $G$ is connected and has exactly $N-1$ edges.
\item $G$ is cycle free and has exactly $N-1$ edges.

\item $G$ is cycle free, but for every two non-adjacent vertices $v,w$, $G \cup \{v,w\}$, contains exactly one cycle.
\end{enumerate}
\end{theorem}
The first studies of trees which can be found in the literature are due to Cayley \cite{cayley_1857,cayley_1875}. 

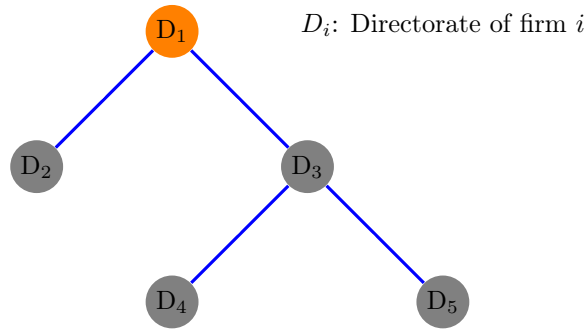
\begin{figure}
\centering
\begin{tikzpicture}[scale=1.8, auto,swap]
    % Draw a 7,11 network
    % First we draw the vertices
            \node[vertex, fill=orange] (x1) at (0,0) {$\mbox{D}_1$};
          
                      \node[vertex, fill=gray] (x3) at (-1,-1) {$\mbox{D}_2$};
            \node[vertex, fill=gray] (x4) at (1,-1) {$\mbox{D}_3$};

            \node[vertex, fill=gray] (x6) at (2,-2) {$\mbox{D}_5$};
            \node[vertex, fill=gray] (x7) at (0,-2) {$\mbox{D}_4$};
         
    % Connect vertices with edges and draw weights

    \begin{scope}[>={Stealth[black]},
              every node/.style={fill=white,circle},
              every edge/.style={draw=blue,very thick}]
    \path [-] (x1) edge node {} (x3);
    \path [-] (x1) edge node {} (x4);      
     \path [-] (x4) edge node {} (x7);  
    \path [-] (x4) edge node {} (x6);

\end{scope}

%\draw [black,dashed] (1,0) ellipse (1.5cm and 0.5cm);
\node[] (x9) at (2,0.05) {$D_i$: Directorate of firm $i$};
    
\end{tikzpicture}
\caption{An ordinary rooted tree representing the relations between the directorates of firms. A connection indicates that two company boards have some common directors. This view allows easily to identify the distance between two directorates $D_i$ of two firms.}\label{fig.otree}
\end{figure}

In Fig. (\ref{fig.otree}) we show an example for an ordinary rooted tree. This ordinary rooted tree represents the relations between the directorates of firms. Specifically, a connection indicates that two company boards are having some common directors that serve on both boards. Due to the fact that such a relation does not imply a natural direction, the connections are undirected. This view allows easily to identify the distance between two directorates $D_i$ of two firms.

\begin{figure}
\centering
\begin{tikzpicture}[scale=1.8, auto,swap]
    % Draw a 7,11 network
    % First we draw the vertices
    \foreach \pos/\name in {{(0,2)/IBM}, {(2,1)/Apple}, {(4,1)/Yahoo},
                            {(0,0)/Cisco}, {(3,0)/Intel}, {(2,-1)/Amazon}, {(4,-1)/HP}}
        \node[vertex, fill=gray] (\name) at \pos {$\mbox{\name}$};
    % Connect vertices with edges and draw weights
    \foreach \source/ \dest /\weight in {Apple/IBM/7, Yahoo/Apple/8,Cisco/Apple/9,
                                         Intel/Apple/7, 
                                         Amazon/Intel/8,
                                         HP/Intel/9}
        \path[edge][green] (\source) -- node[weight] {$$} (\dest);
    
\end{tikzpicture}
\caption{Minimum spanning tree of stocks. }\label{fig.trees}
\end{figure}
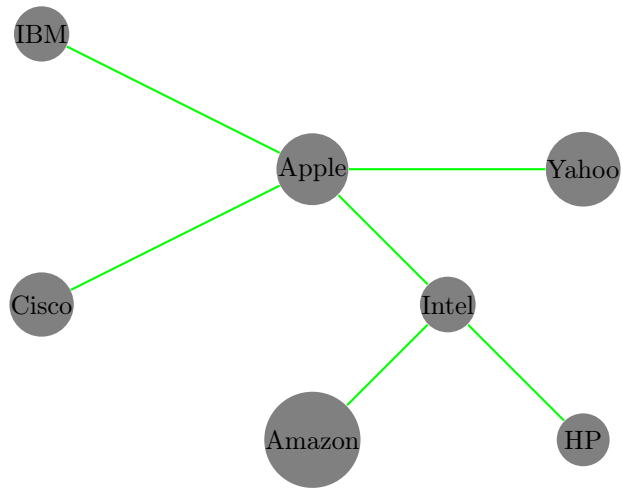

In contrast, in Fig.~(\ref{fig.trees}) we show an example of trees connecting stocks. We want to emphasize that a tree does not possess a hierarchy that means, there is no `top' or `bottom' in the graph of a tree. For this reason the trees shown in Fig.~(\ref{fig.trees}) can be rearranged arbitrarily.
In contrast, rooted trees have a root that is a distinct vertex where all paths point away from it \cite{harary}.

\subsection{Generalized trees}\label{sec_GT}

In this section, we introduce an important extension of trees called {\it generalized trees} (GTs) \cite{dehmer_phd,mehler_2005}. 
Here we only introduce undirected generalized trees. The idea of introducing directed generalized trees have firstly been raised 
in \cite{mehler1}.  When introducing generalized trees, we claim that they are hierarchical and they possess a distinct vertex called root usually present in 
ordinary rooted trees. Besides the edges of an ordinary rooted tree, a GT has more edge types and  
this leads a richer connectivity among vertices. 

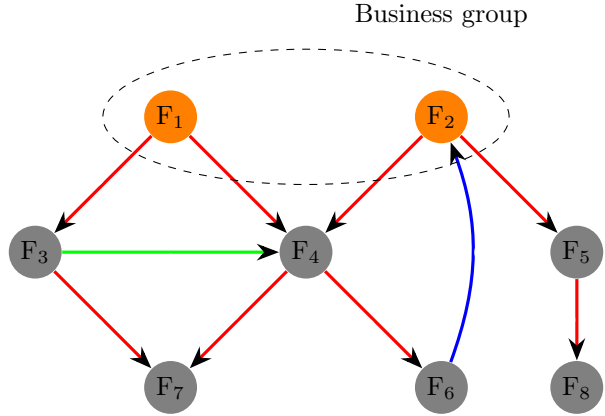
\begin{figure}
\centering
\begin{tikzpicture}[scale=1.8, auto,swap]
    % Draw a 7,11 network
    % First we draw the vertices
            \node[vertex, fill=orange] (x1) at (0,0) {$\mbox{F}_1$};
            \node[vertex, fill=orange] (x2) at (2,0) {$\mbox{F}_2$};
                        
            \node[vertex, fill=gray] (x3) at (-1,-1) {$\mbox{F}_3$};
            \node[vertex, fill=gray] (x4) at (1,-1) {$\mbox{F}_4$};
            \node[vertex, fill=gray] (x5) at (3,-1) {$\mbox{F}_5$};
            
            \node[vertex, fill=gray] (x6) at (2,-2) {$\mbox{F}_6$};
            \node[vertex, fill=gray] (x7) at (0,-2) {$\mbox{F}_7$};
            \node[vertex, fill=gray] (x8) at (3,-2) {$\mbox{F}_8$};
    % Connect vertices with edges and draw weights

    \begin{scope}[>={Stealth[black]},
              every node/.style={fill=white,circle},
              every edge/.style={draw=red,very thick}]
    \path [->] (x1) edge node {} (x3);
    \path [->] (x1) edge node {} (x4);
    \path [->] (x2) edge node {} (x4);
    \path [->] (x2) edge node {} (x5);
    \path [->] (x3) edge[draw=green] node {} (x4);
    \path [->] (x3) edge node {} (x7);
    \path [->] (x4) edge node {} (x7);
    \path [->] (x5) edge node {} (x8);
    \path [->] (x4) edge node {} (x6);
    
    \path [->] (x6) edge[bend right=20, draw=blue] node {$$} (x2); 
   
\end{scope}

\draw [black,dashed] (1,0) ellipse (1.5cm and 0.5cm);
\node[] (x9) at (2,0.75) {Business group};
    
\end{tikzpicture}
\caption{Two generalized trees representing an investment network between firms. Here the firms $F_1$ and $F_2$ are forming a business group. 
The direction of an edge indicates the directionality of the investment. Kernel edges are in red, across edges in green and jump edges are in blue.}\label{fig.gentree}
\end{figure}

\begin{definition}[Generalized Tree]\label{def_gt}
A generalized tree $GT_i$ is defined by a vertex set $V$, an edge set $E$, a level set $L$ and a multi-level function $\mathcal{L}_i$. 
The vertex and edge set define the adjacencies and the level set and the multi-level function induces a hierarchy between the vertices of $GT_i$. The index $i \in V$ indicates the root.
\end{definition}
The multi-level function is defined as follows.
\begin{definition}[Multi-level function]\label{def_mulf}
The function $\mathcal{L}_i: V \rightarrow L \in \mathds{N}$ is called multi-level function with $L_i(i) = 0$.
\end{definition}
The multi-level function $\mathcal{L}_i$ assigns to all vertices an element $l \in L$ that corresponds to the level it will be assigned. The index $i$ refers to the root node, which is assigned to level $l=0$. From these definitions it is immediately clear that a generalized tree is similar to a graph but additionally equipped with a level set $L$ and a multi-level function $\mathcal{L}_i$ introducing a vertex grouping corresponding to the introduction of a hierarchy between vertices and sets thereof.
\begin{definition}[Edge types]
A generalized tree  $GT_i$ has three edges types:
\begin{itemize}
\item Edges with $|\mathcal{L}_i(m) - \mathcal{L}_i(n)|=1$ are called kernel edges ($E_1$).
\item Edges with $|\mathcal{L}_i(m) - \mathcal{L}_i(n)|=0$ are called across edges ($E_2$).
\item Edges with $|\mathcal{L}_i(m) - \mathcal{L}_i(n)|>1$ are called jump edges ($E_3$).
\end{itemize}
Here $m,n \in V$.
\end{definition}
Figure \ref{fig.gentree} shows a generalized tree. The edge types are highlighted by color; kernel edges forming the hierarchy are red, cross edges which do not overjump a level are green and jump edges are blue. Here it is important to emphasize that the two orange nodes representing two firms $F_1$ and $F_2$ are combined into one node representing a business group. That means the shown generalized tree has only one root node. Furthermore, we note that a 
generalized tree is a tree-like graph which may possess cycles \cite{dehmer_GT_MLMTA07}. 
However, a usual graph containing cycles is not hierarchical \cite{harary}.

If one does not collapse the two firms $F_1$ and $F_2$ into a business group but leaves them   as individual nodes, the graph structure shown in Figure \ref{fig.gentree} forms a more complex structure than a generalized tree because it has two root nodes. Such a tree structure has been termed \emph{universal graph}, see \cite{em_j15}.

\subsection{Bipartite networks}

Another network structure that is important to represent economic networks is a bipartite network. A bipartite network consists of two types of nodes. Let's call the first node type $U$ and the second node type $V$.  Edges can only occur between nodes of different type, i.e., 
\begin{eqnarray}
E_{ij}=1 \mbox{ if } v_i\in U \mbox{ and } v_j \in V
\end{eqnarray}
In order to distinguish such a network one writes often $G=(U, V, E)$. In the case of $|U|=|V|$ the network is called a balanced bipartite network. If the connections $E_{ij}$ carry a weight, the graph is called a weighted bipartite network.

\begin{figure}[t!]
\centering
\includegraphics[scale=0.6]{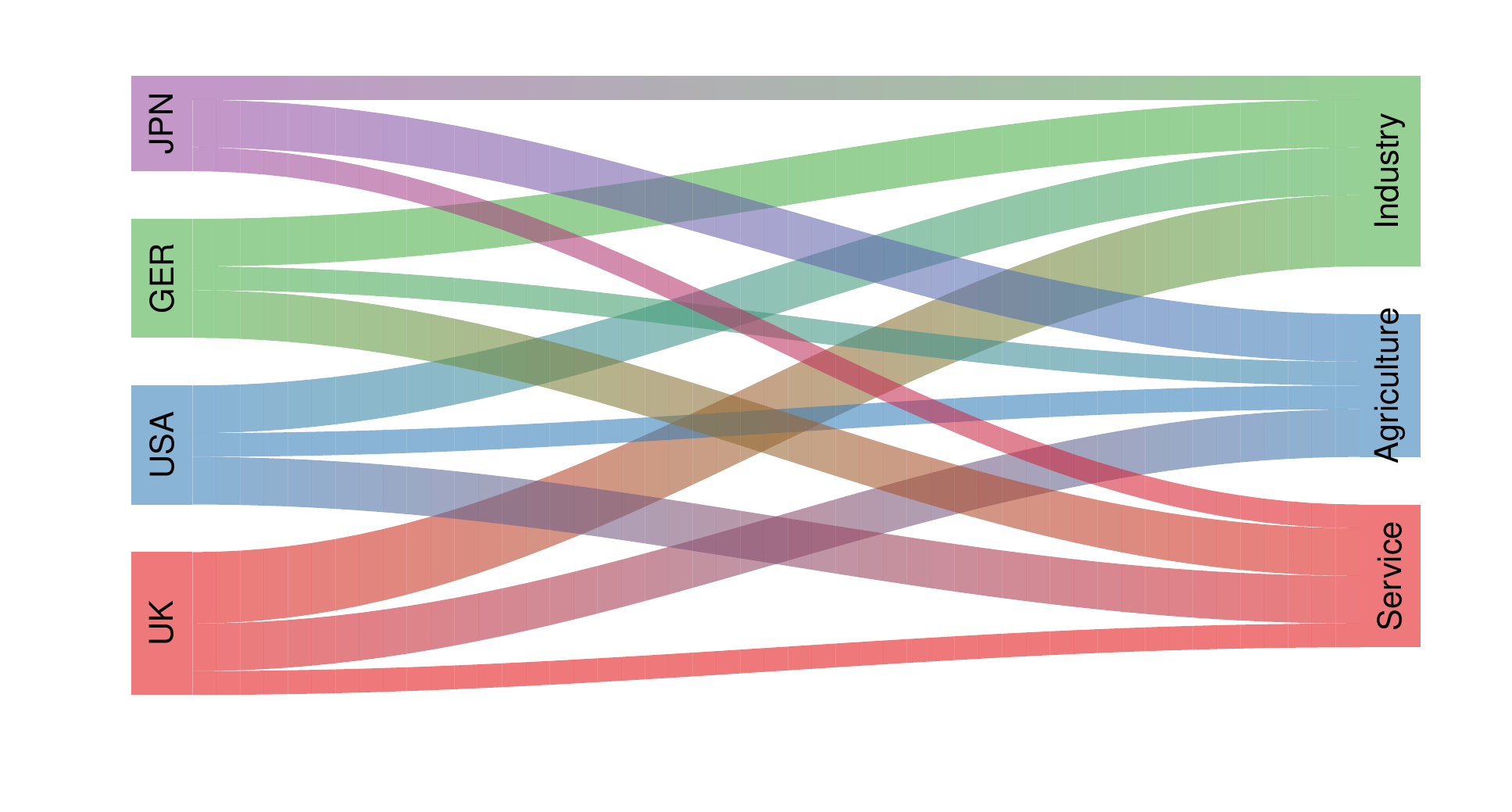}
\caption{Bipartite networks connecting countries with economic sectors. The width of the links is proportional to the strength of the economic sector.}
\label{fig:bip}
\end{figure}

In Fig. \ref{fig:bip} we show an example of a weighted bipartite network. This network connects four countries (UK, USA, GER and JPN) to three economic sectors (industry, agriculture and service). The width of the links is proportional to the strength of the corresponding economic sector. For instance, in this way one can express the contribution of different economic sectors to the GDP of a country or the number of people that are working in the corresponding fields. 

We would like to note that there are many different graphical ways to visualize bipartite networks and in most cases, as seen in Fig. \ref{fig:bip}, the nodes of the bipartite networks are not shown as 'circles', instead, the visualization is more artistic. Nevertheless, one should not forget that the underlying graph is defined in terms of graph theory with a strict meaning.

\subsection{Complex network topologies}

Toward the end of the 1990s two new types of networks have been added to the literature, namely {\it small-world networks} \cite{watts2} and {\it scale-free networks} \cite{AlbertBara_2002rmp}.

Specifically, Watts and Strogatz \cite{watts2}  found that networks, which have been generated according to some rules, have a high clustering coefficient, like regular networks, but also (in average) short distance among vertices, similar to random networks. Hence, these networks, which are called {\it small-world} networks, combine different features from different network classes. With respect to biological networks, small-world networks have been found in, e.g., coexpression, protein and metabolic networks \cite{vannoot_2004,wagner_2001,wilhelm_2003}.

An economic network with small-world characteristics is the directorate interlock network \cite{davis2003small}. The authors study several hundred US firms and banks and thousands of directors on their board of directors during the 1980s and 1990s. They found that the director network where nodes correspond to directors and links correspond to common board positions of two directors is consistent with high clustering coefficients and low average path lengths.

Complementary to this, Barab{\'a}si and Alber found that many real world networks show a scale-free behavior of the node degrees \cite{AlbertBara_2002rmp},
\begin{eqnarray}
P(k) \sim k^{- \gamma}.
\end{eqnarray}
To explain this common feature  Barab{\'a}si and Albert introduced a model \cite{barabasi_1999} now known as Barab{\'a}si-Albert (BA) or {\it preferential attachment} model \cite{newman_2003} that results in so called {\it scale-free} networks which have a degree distribution following a power law \cite{barabasi_1999}. A major difference between the {\it preferential attachment} model and the other algorithms described above to generate random or small-world networks is that the BA model does not assume a fixed number of vertices $N$ and then rewires them iteratively with a fixed probability, but in this model $N$ grows. Each newly added vertex is connected with a certain probability (which is not constant) to other vertices already present in the network. This attachment probability,
\begin{eqnarray}
p_i = \frac{k_i}{\sum_j k_j},
\end{eqnarray}
 is proportional to the degree, $k_j$, of these vertices, explaining the name of the model. This way, each new vertex is added to $e \in \mathbb{N}$ existing vertices in the network.

For instance, Garlaschelli et al. found that the network of investment markets follows a scale-free distribution \cite{garlaschelli2005scale}. Further examples of economic networks exhibiting a scale-free degree distribution, e.g., for interbank networks or world trade networks, can be found in \cite{boss_2004,SOUMA2003396,arnold2006topology,serrano2003topology}.

\section{Future directions and discussion}\label{future}

A natural prerequisite for any economic network analysis is the creation or inference of networks.  For this reason, we think it would be useful to conduct some comparative analyses to identify the best method for different types of economic networks. In order to perform such  analyses one would need to define what in this context 'best' means. The problem is that in most cases the true economic network is not known. For instance, the true financial network of the NYSE connecting stocks is unknown. For this reason, one needs to define context-specific measures that allow, potentially in an indirect way, the assessment of the inferred network structure. In the case of financial networks this could be accomplished by utilizing financial networks in predictive forecasting models to estimate, e.g., future stock prices. That means one can compare predictive models taking the network structure into consideration vs. predictive models that do not. Better prediction results in the former case could indicate that the inferred financial network captures actually meaningful information that can be translated into beneficial forecasts. 

As a further particular example we want to look at investor networks \cite{macleod_2009}. MacLeod \cite{macleod_2009} investigated social properties
of stakeholders by using these networks. Also, they compared them with so-called advocacy networks of social activists. It's worth mentioning that these network classes  could be compared by comparative graph measures \cite{em_jJuly020115}. This would extent the work in \cite{macleod_2009} considerably as it leads to a thorough graph-theoretical treatment of this problem. Another example of analyzing investor networks by using quantitative network measures can be found in 
\cite{ozsoylev_2014}.

Research on networks in economics and finance has exploded for many reasons. Financial and economic systems are often complex and it is necessary to pay attention to the patterns of interactions to understand the behavior of systems and agents therein. Increasingly available data from various sources for different systems enables researchers to test and apply theories \cite{jackson1996strategic,kirman1997economy,ozsoylev_2014,walden2017trading,hagstromer2016network}, but also conduct explanatory research \cite{onnela2003dynamics,onnela2003dynamicBlack,boss2004network,iori2008network,fagiolo2009world,tumminello2012identification,baltakys2017multilayer} for better empirical understanding of financial and economic systems. At the same time, network research in economics and finance is quite fragmented. This research is not only fragmented in terms of research topics and applications, but also in terms of research approaches, academic disciplines and their journals.
 
In economics, there have been several application areas of networks, including network games \cite{ballester2006s,bramoulle2015games}, labor markets \cite{calvo2007networks,beaman2016social}, international trade \cite{rauch1999networks,chaney2016networks} and social networks overall  \cite{morelli2017empathy}. This type of research is typically published in well-established journals in economics, but especially research on international trade has also published in cross-disciplinary journals (see, for example \cite{bhattacharya2008international,jackson2015networks,saracco2016detecting}).
 
In finance, complex systems and networks offer potential for better analysis and monitoring of research in systemic risk in financial systems, and research on this topic has been published in both financial journals \cite{acemoglu2015systemic,markose2012too,hautsch2014financial,diebold2014network,billio2012econometric} and multi-disciplinary journals \cite{haldane2011systemic,battiston2012liaisons,cimini2015systemic,battiston2016complexity}. As emphasized in \cite{jackson2015past}, in the area of systemic risk, the gap between theory and applications still needs to be closed, which is important because in this area network theory can have an immediate as well as lasting impact. Secondly, research papers on networks in the financial markets have mainly been published in {\em multi-disciplinary, complexity, and physics} journals rather than in finance (see, for example, \cite{onnela2003dynamics,onnela2003dynamicBlack,emmert2010influence,emmert2010identifying,battiston2012debtrank,tumminello2012identification,musciotto2016patterns,gualdi2016statistically}), though exceptions recent exist \cite{han2013social,ozsoylev_2014,ahern2017information}. Network methods can be used to investigate investors joint behavior and interaction, and given that investor network structure is important for the stock price dynamics \cite{walden2017trading}, it is quite surprising that finance journals have so slowly and marginally adopted network methods in the research of investor behavior.
 
In econophysics, papers are often data-driven and exploratory whereas papers published in finance journals rely on models, typically under the neoclassical economics paradigm of rational individual choice. The research questions and approaches can be very different between different journals, indeed, which can partially be explained by researchers' different interests and backgrounds. Nevertheless, different research approaches (e.g. exploratory vs confirmatory research) can benefit from each other and one of the most important possibilities in network research in finance is to have actual interaction between research published in different journal categories, i.e. multidisciplinary journals, complexity, (econo)physics, and finance journals. Particularly, data-driven research can feed 'traditional' finance research by raising observations that should be theoretically explained, and, on the other hand theoretical models should be carefully reflected and evaluated by large data sets with alternative methods used to use in non-financial journals. Also, there would be possibilities of using methods developed in exploratory research in financial modeling. For example, \cite{tumminello2012identification} provide a method to estimate links between investors and detect investor communities, which could be used to verify financial theories in the interaction of individual investors and information transformation. For example, as there is a theoretical relation between investor networks and volatility dynamics \cite{walden2017trading}, methods developed to estimate links between investor \cite{tumminello2012identification,baltakys2017multilayer} could be used to produce state variables to be augmented into volatility models for better risk management and option pricing accuracy. Addressing this question requires experience on both network inference methods, which are typically published in multi-disciplinary journals (see, for example \cite{tumminello2012identification}), as well as established time-series models.

\section{Conclusion}\label{con}

The purpose of this survey was to showcase the use of graph theoretic methods for studying economic problems. We hope this will help in making the study of economic networks more popular and accessible in the economics and finance literature because of the tremendous potential of such approaches for shedding light on our global, interconnected world.

\section{Acknowledgments}

Matthias Dehmer thanks the Austrian Science Funds for supporting this work (project P26142). JK received funding from the European Union Horizon 2020 research and innovation
programme under the Marie Sklodowska-Curie grant agreement No 675044 ``BigDataFinance''.

\bibliographystyle{plain}

%\bibliography{matthias_bibtex_current,future_directions}

%\bibliography{/Users/fes/fes/Artikel/Bib_ref/ref_list,/Users/fes/Local/Mails/CV/Bibref/bib_ref_journals,/Users/fes/Local/Mails/CV/Bibref/bib_ref_books_editor,/Users/fes/Local/Mails/CV/Bibref/bib_ref_book_chapters,matthias_bibtex_current,future_directions,juho}

\end{document}